\documentclass[runningheads]{llncs}
\usepackage[T1]{fontenc}
\usepackage{xcolor}
\usepackage{subcaption}
\usepackage{amsmath}
\usepackage{amssymb}  
\usepackage{hyperref}
\usepackage{makecell}
\usepackage{adjustbox}
\usepackage{multirow}
\usepackage{pifont}
\usepackage{enumitem}

\usepackage{graphicx,verbatim}
\begin{document}
\title{Reconsidering Explicit Longitudinal Mammography Alignment for Enhanced Breast Cancer Risk Prediction}

\titlerunning{Reconsidering Explicit Alignment for Breast
Cancer Risk Prediction}
%
\author{Solveig Thrun\inst{1}\  \and
Stine Hansen\inst{2} \and
Zijun Sun\inst{3} \and Nele Blum\inst{4} \and  Suaiba A. Salahuddin\inst{1} 
\and  Kristoffer Wickstrøm\inst{1} \and Elisabeth Wetzer\inst{1} \and Robert Jenssen\inst{1,5,6} \and Maik Stille\inst{4} \and Michael Kampffmeyer\inst{1,5} 
}
\authorrunning{S. Thrun et al.}
%
\institute{Department of Physics and
Technology, UiT The Arctic University of Norway, Tromsø, Norway 
\email{solveig.thrun@uit.no}\\ \and SPKI The Norwegian Centre for Clinical Artificial Intelligence, University Hospital of North Norway, Tromsø, Norway \and Department of Computer Science and Engineering, University of Bologna, Bologna, Italy  \and Fraunhofer Research Institution for Individualized and Cell-Based Medical Engineering IMTE, Lübeck, Germany \\ \and
Norwegian Computing Center, Oslo, Norway \and
Pioneer Centre for AI, University of Copenhagen, Copenhagen, Denmark }
    
\maketitle              
\begin{abstract}
Regular mammography screening is essential for early breast cancer detection. Deep learning-based risk prediction methods have sparked interest to adjust screening intervals for high-risk groups. While early methods focused only on current mammograms, recent approaches leverage the temporal aspect of screenings to track breast tissue changes over time, requiring spatial alignment across different time points. Two main strategies for this have emerged: explicit feature alignment through deformable registration and implicit learned alignment using techniques like transformers, with the former providing more control. However, the optimal approach for explicit alignment in mammography remains underexplored. In this study, we  provide insights into where explicit alignment should occur (input space vs. representation space) and if alignment and risk prediction should be jointly optimized. We demonstrate that jointly learning explicit alignment in representation space while optimizing risk estimation performance, as done in the current state-of-the-art approach, results in a trade-off between alignment quality and predictive performance and show that image-level alignment is superior to representation-level alignment, leading to better deformation field quality and enhanced risk prediction accuracy. The code is available at
\url{https://github.com/sot176/Longitudinal_Mammogram_Alignment.git}

\end{abstract}
\section{Introduction}
Breast cancer is among the most prevalent cancers globally~\cite{cancerstat} and is primarily detected through regular mammography screening, capturing cranio-caudal (CC) and medio-lateral oblique (MLO) views~\cite{pharmaceutics13050723}. However, screening frequency is limited by healthcare capacity, creating a trade-off between early detection and resource availability~\cite{mammoscreen}.
 A promising solution is developing more accurate risk prediction models for personalized screening, optimizing follow-up intervals based on individual risk. Although current mammograms can be used for this purpose, radiologists typically compare them with prior mammograms to distinguish benign from malignant changes, enhancing diagnostic accuracy~\cite{akwo2024access,prioirmammo}.\\
\indent Similarly, recent deep learning-based studies have demonstrated that longitudinal mammography has the potential to improve risk prediction, surpassing models based solely on current images~\cite{miccai2023,Wang2023,dadsetan2022deep,Kar_Longitudinal_MICCAI2024,Wan_Ordinal_MICCAI2024}, emphasizing the importance of analyzing temporal tissue changes. However, effective use of longitudinal mammograms requires precise spatial alignment or registration, as misalignment can obscure abnormalities and distort risk assessments. This is challenging due to the non-rigid nature of breast tissue~\cite{guo2006breast} and variations in breast compression during examinations. \\
\indent While various alignment strategies exist, direct comparisons are scarce, highlighting the need for deeper insights. Alignment methods fall into two categories: explicit~\cite{Wan_Ordinal_MICCAI2024,loizidou2022automatic,dadsetan2022deep} and implicit~\cite{miccai2023,Kar_Longitudinal_MICCAI2024}. For longitudinal mammography, explicit alignment is preferred as it directly corrects spatial misalignment, ensuring anatomical consistency over time \cite{Chen2023}. 
This approach preserves crucial information, reduces artifacts, and aids in tracking tissue changes. Therefore, we focus on explicit alignment, in form of affine or deformable registration in either the feature space or input space. Early methods performed registration in the input space before feature extraction using conventional registration techniques~\cite{loizidou2022automatic,dadsetan2022deep}, which is computationally expensive and time-consuming, as it typically requires per-image optimization. Recent state-of-the-art approaches therefore align feature maps directly in the feature space~\cite{Wan_Ordinal_MICCAI2024}. However, this can degrade performance due to the large receptive field of the encoders and downsampling. Due to the absence of direct comparisons of alignment mechanisms, there is a lack of guidance on explicit alignment strategies for longitudinal mammography.\\
\indent  To address the lack of insights into the choice of alignment, this study aims to evaluate explicit alignment strategies for longitudinal mammography and their impact on risk prediction. We examine whether alignment should occur at the image or feature level, demonstrating that feature alignment may not be optimal despite its use in current state-of-the-art approaches. We further demonstrate a clear objective function mismatch for feature-level approaches, where the risk prediction and quality alignment objectives typically disagree. Further, to facilitate efficient image level alignment, we take inspiration from recent advances in registration and transfer them to the mammography domain, introducing MammoRegNet, which to our knowledge is the first deep learning model for high-quality longitudinal mammogram registration.\\
Our \textbf{main contributions} are:
\begin{enumerate}[noitemsep,  topsep=-1pt]   
\item We conduct a comprehensive evaluation of explicit alignment strategies and their effect on risk prediction performance using the publicly available EMBED~\cite{embed} and CSAW-CC~\cite{csawcc} datasets.
    \item We provide insights into the trade-offs between feature-level alignment quality and risk prediction performance.
    \item We demonstrate that image-level alignment enhances both deformation field quality and risk prediction performance.
    \item We introduce MammoRegNet, the first deep learning-based image registration model for aligning longitudinal mammography images.
\end{enumerate}

\section{Methods}
\begin{figure}[tbp]
    \centering  
    \begin{subfigure}{0.2\textwidth}
            \raggedright  
        \includegraphics[height=2.2cm]{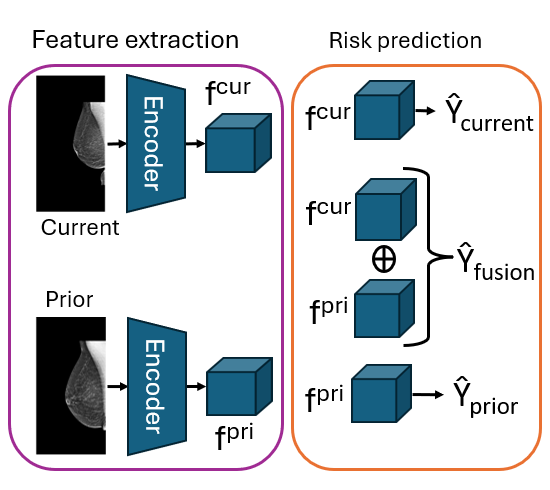}
        \caption{NoAlign}
        \label{fig:noalign}
    \end{subfigure}
    \begin{subfigure}{0.2\textwidth}
        \centering
    \includegraphics[height=2.2cm]{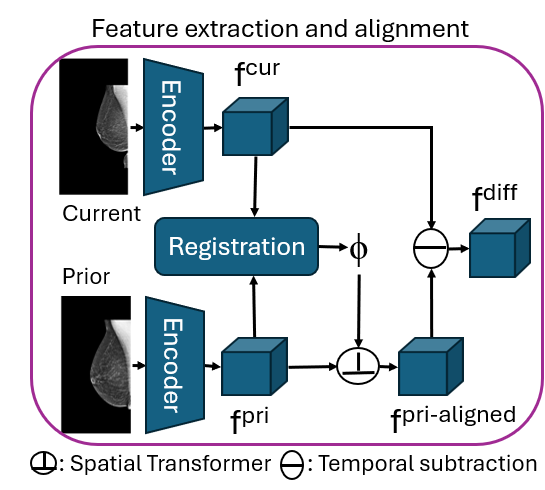}
        \caption{FeatAlign}
                \label{fig:featalign}
    \end{subfigure}
    \begin{subfigure}{0.21\textwidth}
            \raggedright 
    \includegraphics[height=2.2cm]{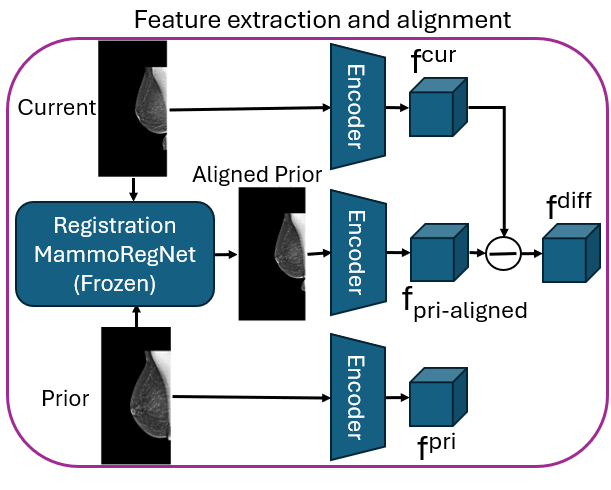}
        \caption{ImgAlign}
                \label{fig:imgalign}
    \end{subfigure}
     \begin{subfigure}{0.21\textwidth}
         \raggedleft   
    \includegraphics[height=2.2cm]{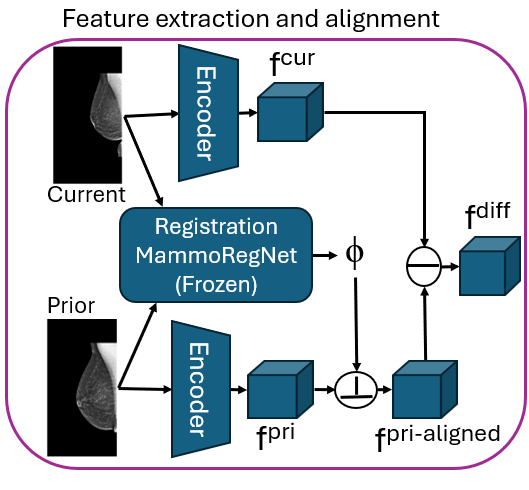}
        \caption{ImgFeatAlign}
                \label{fig:imgfeatalign}
    \end{subfigure}
     \begin{subfigure}{0.12\textwidth}
        \centering
\includegraphics[height=2.25cm]{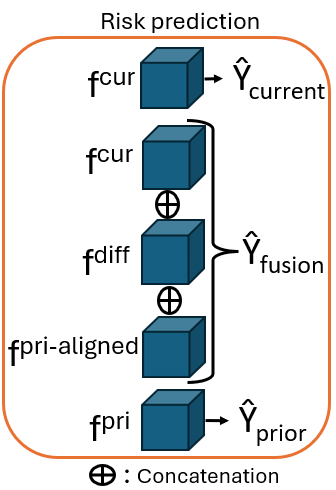}
      \caption{Risk }
    \label{fig:alignrisk}
       \end{subfigure}
    \caption{Overview of longitudinal risk prediction methods: (a) Direct feature extraction without alignment, (b) Feature-level alignment, (c)  Image-level alignment with MammoRegNet, (d) Applying MammoRegNet’s deformation field in feature space, and (e) Risk prediction using alignment methods (b), (c), and (d).}
    \label{fig:alignmentmethods}
\end{figure}

Risk prediction estimates the probability of developing breast cancer within a specified time frame, focusing on a 5-year period following a specific mammogram. This process requires the alignment of longitudinal mammography images. Given a current image \(\mathbf{I}^{\text{cur}} \in \mathbb{R}^{H \times W}\) or a feature map \(\mathbf{f}^{\text{cur}} \in \mathbb{R}^{C \times H \times W}\) and a prior image \(\mathbf{I}^{\text{pri}} \in \mathbb{R}^{H \times W}\) or a feature map \(\mathbf{f}^{\text{pri}} \in \mathbb{R}^{C \times H \times W}\), the goal of explicit alignment is to determine a deformation field \(\boldsymbol{\phi} \in \mathbb{R}^{2 \times H \times W}\) that maximizes similarity between the transformed prior and current images (or feature maps). Effective alignment is crucial for precise anatomical registration and a smooth deformation field. To assess its impact on risk prediction, we compare different alignment strategies within a consistent framework: Fig.~\ref{fig:alignmentmethods}: (a) No explicit alignment (Sec. 2.1), (b) feature-level alignment (Sec. 2.2), as done in the current state-of-the-art approach~\cite{Wan_Ordinal_MICCAI2024}, enhances detection of subtle patterns by focusing on high-level semantics, though it may miss important spatial relationships, and (c) image-level alignment (Sec. 2.3), retaining spatial and textural details crucial for identifying subtle risk indicators. This study examines how image-level and feature-level alignment strategies affect risk prediction in mammography.

\subsection{Risk Prediction Baseline}
Our risk prediction baseline (Fig.~\ref{fig:noalign}) builds on~\cite{mirai} and~\cite{Wan_Ordinal_MICCAI2024}, incorporating multilevel joint learning and a cumulative probability layer for time-dependent risk estimation. It uses an image encoder to generate two feature maps: current \(\mathbf{f}^{\text{cur}}\) and prior \(\mathbf{f}^{\text{pri}}\). The model predicts breast cancer risk at three levels: (1) Fused Feature Prediction, using concatenated unaligned prior and current features; (2) Current Feature Prediction, using only current features; and (3) Prior Feature Prediction, using unaligned prior features. Following~\cite{Wan_Ordinal_MICCAI2024}, each level outputs a six-dimensional risk vector: the first five elements estimate cancer occurrence over the next five years, and the last indicates the probability of remaining cancer-free. Only fused feature predictions determine the final risk assessment. The risk loss function employs binary cross-entropy at each prediction level, with masking to focus on observed time points and ignore unobserved future data.

\subsection{Feature-Level Alignment}
To evaluate feature-level alignment, we follow the current state-of-the-art approach~\cite{Wan_Ordinal_MICCAI2024}, using two $3\times3$ convolutional layers, batch normalization~\cite{bn}, and ReLU activation~\cite{relu}. This block processes the concatenated feature maps to estimate a deformation field \(\boldsymbol{\phi}\). A spatial transformer block applies this deformation to align the prior feature maps \(\mathbf{f}^{\text{pri-aligned}}\). The alignment is trained using $\mathcal{L}_{\text{feat}}$. To further encourage smooth alignment, as commonly done in image-level alignment approaches, we further constrain their loss with a (optional) regularization of the displacement field and a Jacobian Determinant (JD) loss to penalize negative Jacobian Determinants resulting in
\begin{equation}\label{lossfeat}
    \mathcal{L}_{\text{feat}} = \alpha \left( \left\| \mathbf{f}^{\text{pri-aligned}} - \mathbf{f}^{\text{cur}} \right\|_2^2 \right) + \beta \left( \sum_{\mathbf{p} \in \Omega} \left\| \nabla \phi(\mathbf{p}) \right\|^2 + \lambda \text{JD} (\phi) \right)\, ,
\end{equation}
where $\alpha$ and $\beta$ are weighting factors, $\Omega$ represents the set of spatial locations $\mathbf{p}$, $\nabla$ denotes the spatial derivative of $\phi$ at $\mathbf{p}$, and $\lambda$  is the regularization parameter. Fig.~\ref{fig:imgalign} illustrates feature-level alignment, while Fig.~\ref{fig:alignrisk} shows risk prediction using the aligned feature maps. This approach enhances the risk prediction baseline by incorporating four key feature maps, as in~\cite{Wan_Ordinal_MICCAI2024}:  current \(\mathbf{f}^{\text{cur}}\), prior \(\mathbf{f}^{\text{pri}}\), aligned prior \(\mathbf{f}^{\text{pri-aligned}}\), and difference features \(\mathbf{f}^{\text{diff}} = \mathbf{f}^{\text{cur}} - \mathbf{f}^{\text{pri-aligned}}\), which capture temporal changes. Following \cite{mirai}, positional encoding is applied to \(\mathbf{f}^{\text{diff}}\) to account for time gaps between screenings. While Current and Prior Feature predictions remain unchanged from the risk prediction baseline, Fused Feature Prediction integrates aligned prior, current, and difference features.

\subsection{Image-Level Alignment}
Traditional registration methods for longitudinal mammograms~\cite{loizidou2022automatic,dadsetan2022deep} are often computationally expensive, requiring per-image optimization, and are less accurate than deep learning approaches. As an alternative to feature-level methods, we develop MammoRegNet, a deep learning-based registration network adapted from the Non-Iterative Coarse-to-Fine Transformer (NICE-Trans) network~\cite{nicetrans}.  Originally designed for 3D brain MRI registration, NICE-Trans was modified for 2D mammography by replacing 3D convolutions with 2D ones and converting the Swin Transformer and Spatial Transformer blocks to 2D. MammoRegNet features an encoder with two parallel paths for current and prior images, each with five convolutional modules. The concatenated features are processed by a decoder with four SwinTrans modules and a convolutional module. Each SwinTrans module includes a convolutional layer and four SwinTrans blocks~\cite{swintransblock}, with patch-expanding blocks to enhance spatial resolution. The decoder performs registration in a coarse-to-fine manner, starting with affine registration followed by four deformable refinements. Following~\cite{nicetrans}, MammoRegNet is trained using 
\begin{equation}\label{losssegunc}
       \mathcal{L}_{\text{img}} =  (1 - \text{NCC}_{\text{affine}}) + (1 - \text{NCC}_{\text{final}}) + \gamma \left( \sum_{\mathbf{p} \in \Omega} \left\| \nabla \phi(\mathbf{p}) \right\|^2 + \lambda \text{JD} (\phi) \right)\, .
\end{equation} 
This unsupervised loss function includes image similarity measured by the normalized cross-correlation (NCC) between the current mammogram and both the affine- and final-transformed prior mammograms and a regularization term. Fig.~\ref{fig:imgalign}  shows the image-level alignment pipeline, with risk prediction in Fig.~\ref{fig:alignrisk}. We also evaluate applying MammoRegNet's deformation field in feature space (Fig.~\ref{fig:imgfeatalign}), with risk prediction outlined in Fig.~\ref{fig:alignrisk}.

\section{Experiments and Results}
\subsection{Experimental Setting}
\subsubsection{Dataset and Pre-processing}
This study uses two large, publicly available mammography datasets to ensure reproducibility and encourage further research. The Emory Breast Imaging Dataset (EMBED)~\cite{embed} comprises 364,564 mammograms from 23,256 patients, collected from 2013 to 2020 using Hologic, GE, and Fujifilm systems, accessible \href{https://aws.amazon.com/marketplace/pp/prodview-unw4li5rkivs2#overview}{here}. The Cohort of Screen-Aged Women Case Control (CSAW-CC)~\cite{csawcc} comprises images from 1,103 breast cancer diagnoses and 7,850 healthy controls, collected from 2008 to 2015 at Karolinska University Hospital, Sweden, available \href{https://snd.se/en/catalogue/dataset/2021-204-1}{here}. Following~\cite{Wan_Ordinal_MICCAI2024}, for the risk prediction dataset we included patients with at least five years of follow-up data for risk prediction. Images were resized to $1024 \times 512$ pixels while maintaining aspect ratio, and were divided into training, validation,
and test sets (5:2:3 ratio).  For registration training, we randomly selected 1,000 patients from the risk dataset, using two images per laterality and view combination to reduce training time and split it into training, validation, and test sets (5:2:3 ratio). \subsubsection{Evaluation Metrics} Following~\cite{nicetrans}, we evaluate registration performance using NCC to assess image-wise similarity and measure deformation field quality for both image-level and feature-level alignment by the percentage of Negative Jacobian Determinants (NJD)~\cite{reg_njd}. Risk prediction performance is assessed using the concordance index (C-index) and Area Under the Curve (AUC) for the 1–5 year risk prediction task, as in~\cite{mirai,miccai2023,Wan_Ordinal_MICCAI2024}. 95\% confidence intervals (CI) for NCC, NJD, AUC, and C-index are estimated via bootstrapping with 1,000 iterations.

\subsubsection{Implementation Details} We use a ResNet-18~\cite{resnet18}, pre-trained on ImageNet~\cite{imagenet}, without the last fully connected layer. For feature-level alignment, risk prediction and alignment are jointly trained using a combined loss with $\alpha = 1e-2$ in Eq.~\ref{lossfeat} as in~\cite{Wan_Ordinal_MICCAI2024}. For image-level alignment, only the risk loss is used, with the registration model frozen. We use the Adam optimizer~\cite{adam} (learning rate = $1e-4$, weight decay = $1e-5$), a batch size of 12, and 100 epochs. Early stopping and learning rate decay are based on the validation C-index: the learning rate halves after 5 stagnant epochs, and training stops after 15. For MammoRegNet, we use the Adam optimizer (learning rate = $1e-4$, weight decay = $1e-6$), a batch size of 20, and 100 epochs, with 1,500 image pairs randomly selected per epoch. $\lambda$ is set to $1e-5$, and $\gamma$ to 1 in Eq.~\ref{losssegunc}. All models were implemented in PyTorch and trained on an NVIDIA GeForce RTX 3090, taking between 14 and 72 hours.

    \begin{table}[tb]
    \centering
     \scriptsize
    \caption{1–5 Year Breast Cancer Risk Prediction of different alignment methods: C-index and AUC with 95\% CI in parentheses.}
    
    \begin{adjustbox}{center}
    \begin{tabular}{|c|c|c|*{5}{c}|} 
    \hline
    & \multirow{2}{*}{} & \multirow{2}{*}{C-index (\%) $\uparrow$} & \multicolumn{5}{c|}{Follow-up year AUC (\%) $\uparrow$}   \\
    \cline{4-8}
    & &  & 1-year & 2-year & 3-year & 4-year & 5-year \\
    \hline
    
    \multirow{3}{*}{\rotatebox{90}{\textbf{EMBED}}}  
      & \makecell{NoAlign} & \makecell{63.6\\(61.2-66.0)} & \makecell{63.9\\(61.0-66.8)} & \makecell{64.1\\(61.5-66.7)} & \makecell{64.1\\(61.7-66.6)} & \makecell{64.7\\(62.1-67.2)} & \makecell{59.8\\(55.8-63.8)} \\
    & \makecell{FeatAlign} & \makecell{64.9\\(62.4-67.3)} & \makecell{64.2\\(61.1-67.2)} & \makecell{64.5\\(61.8-67.1)} & \makecell{65.4\\(62.9-67.7)} & \makecell{65.3\\(62.8-67.7)} & \makecell{\textbf{65.8}\\\textbf{(61.7-69.4)}} \\
    & \makecell{FeatAlignReg} & \makecell{64.1\\(61.6-66.6)} & \makecell{63.9\\(60.8-66.9)} & \makecell{64.8\\(62.0-67.5)} & \makecell{64.6\\(62.1-67.0)} & \makecell{63.5\\(61.0-65.8)} & \makecell{60.6\\(56.7-64.4)} \\
 & \makecell{ImgFeatAlign} &  \makecell{66.1\\(63.6-68.5)}  &  \makecell{65.9\\(63.0-68.6)}  &  \makecell{66.2\\(63.4-68.7)}  &  \makecell{67.0\\(64.5-69.5)}  &  \makecell{65.9\\(63.4-68.5)} & \makecell{64.2\\ (60.3-68.1) } \\ 
    & \makecell{ImgAlign} & \makecell{\textbf{66.9} \\\textbf{(64.6-69.3)}} & \makecell{\textbf{67.4} \\\textbf{(64.4-70.4)}} & \makecell{\textbf{67.5}\\\textbf{(64.9-70.2)}} & \makecell{\textbf{68.1}\\\textbf{(64.9-70.6)}} & \makecell{\textbf{66.6}\\\textbf{(64.0-69.2)}} & \makecell{65.6\\ (61.6-69.4)} \\

    \hline
    
    \multirow{3}{*}{\rotatebox{90}{\textbf{CSAW-CC}}}  
      & \makecell{NoAlign} & \makecell{54.1\\(51.9-56.1)} & \makecell{54.2\\(51.5-56.9)} & \makecell{55.5\\(53.4-57.7)} & \makecell{55.6\\(53.5-57.8)} & \makecell{54.7\\(52.5-56.9)} & \makecell{55.7\\(53.2-58.2)} \\
    & \makecell{FeatAlign} & \makecell{ 57.0\\(54.8-58.9)} & \makecell{57.3\\(54.5-60.1)} & \makecell{58.2\\(55.8-60.2)} & \makecell{58.5\\(56.2-60.6)} & \makecell{ 58.5\\(56.2-60.6)} & \makecell{60.5\\(58.0-63.0)} \\
    & \makecell{FeatAlignReg} & \makecell{56.0\\(53.8-57.9)} & \makecell{56.3\\(53.5-59.0)} & \makecell{57.3\\(55.3-59.5)} & \makecell{57.8\\(55.7-60.1)} & \makecell{57.6\\(55.5-59.9)} & \makecell{60.4\\(57.8-62.9)} \\
 & \makecell{ImgFeatAlign} &  \makecell{60.6\\(58.7-62.5)}  &  \makecell{ 61.8\\(59.2-64.5)}  &  \makecell{61.9\\(59.8-63.9)}  &  \makecell{ 62.7\\(60.7-64.8)}  &  \makecell{ 61.3\\(59.2-63.4)} & \makecell{63.9\\ (61.4-66.5) } \\ 
    & \makecell{ImgAlign} & \makecell{\textbf{61.3}\\\textbf{(59.3-63.4)}} & \makecell{ \textbf{62.1}\\\textbf{(59.3-64.9)}} & \makecell{ \textbf{62.4}\\\textbf{(60.4-64.4)}} & \makecell{ \textbf{63.8}\\\textbf{(61.7-65.9)}} & \makecell{\textbf{64.0}\\\textbf{(61.9-66.1)}} & \makecell{\textbf{66.2}\\\textbf{(63.7-68.7)}} \\

    \hline
    \end{tabular}
    \end{adjustbox}

    \label{tab:riskpred}
\end{table} 

\begin{figure}[tb]
     \centering
\includegraphics[width=0.8\textwidth]{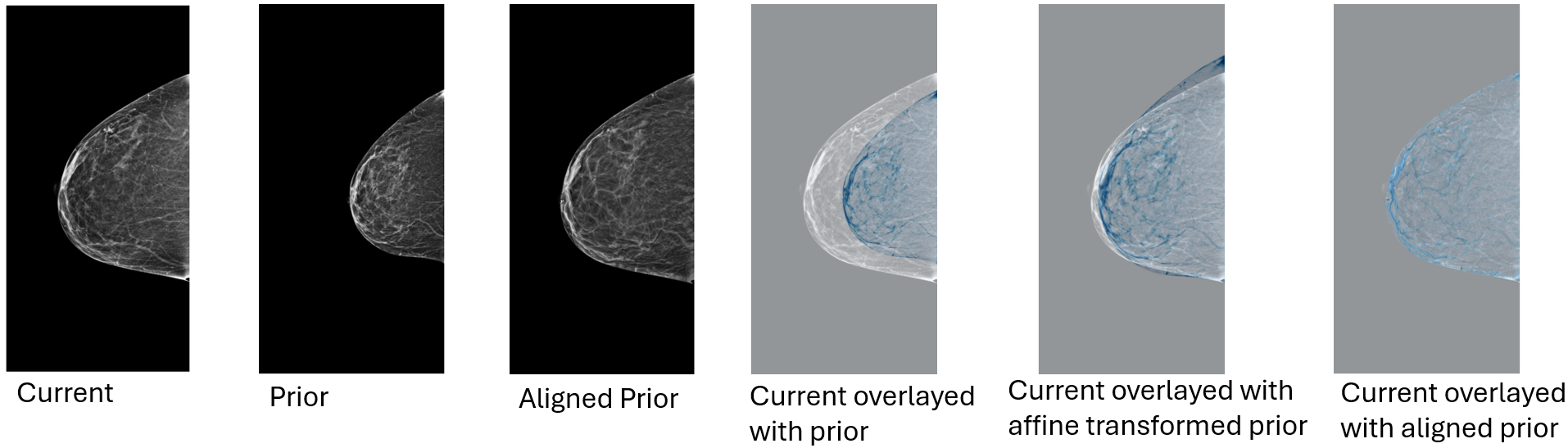}
  \caption{Registration result of MammoRegNet.   }
     \label{fig:regmodel}
 \end{figure}

\subsection{Results}
\subsubsection{Effect of Feature- vs. Image-Level Alignment on Risk Prediction}
We assess the impact of feature-level and image-level alignment on risk prediction performance (Table~\ref{tab:riskpred}) by comparing NoAlign (Fig.~\ref{fig:noalign} ), FeatAlign (Fig.~\ref{fig:featalign} ) with $\beta=0$, FeatRegAlign (Fig.~\ref{fig:featalign}) with $\beta=0.2$, ImgFeatAlign (Fig.~\ref{fig:imgfeatalign}), and ImgAlign (Fig.~\ref{fig:imgalign}).  Our results demonstrate that image-level deformation enhances risk prediction, with ImgFeatAlign outperforming feature-level methods and ImgAlign achieving the highest accuracy. Both datasets show consistent results, suggesting that aligning images before feature extraction better preserves context and enhances feature representativeness, underscoring the benefits of image-level alignment for robust risk prediction.  Table~\ref{tab:mammoregnet} presents the performance of MammoRegNet and Fig.~\ref{fig:regmodel} shows a registration result. MammoRegNet achieves precise alignment across datasets and deformation scales, effectively handling both large and small deformations, thereby enhancing the accuracy and reliability of longitudinal analyses by correcting spatial inconsistencies.

\begin{table}[t]
    \centering
     \scriptsize
    \caption{MammoRegNet registration performance (95\% CI in parentheses).}
    \begin{adjustbox}{center}
    \begin{tabular}{|c|c|c|c|c|}
    \hline
     &  NCC before (\%) $\uparrow$ & NCC affine (\%) $\uparrow$ & NCC final (\%)  $\uparrow$ &  NJD (\%) $\downarrow$ \\
    \hline
    \makecell{EMBED} &  \makecell{74.5} & \makecell{83.1\\(82.8-83.4)} & \makecell{95.5\\(95.5-95.6)} & \makecell{ 0.0016 \\(0.0013-0.0020)}   \\
 \makecell{CSAW-CC} &  \makecell{78.3} & \makecell{ 85.2\\(84.8-85.7)} & \makecell{ 95.8\\(95.5-96.0)} & \makecell{ 0.0129\\(0.0088-0.0180)}   \\
    \hline
    \end{tabular}
    \end{adjustbox}
    \label{tab:mammoregnet}
\end{table}

\subsubsection{Comparison of Deformation Field Quality}
\begin{figure}[t]
    \centering
    \begin{subfigure}{0.32\textwidth}
        \centering
        \begin{subfigure}{0.45\textwidth}
            \centering
            \includegraphics[width=\textwidth]{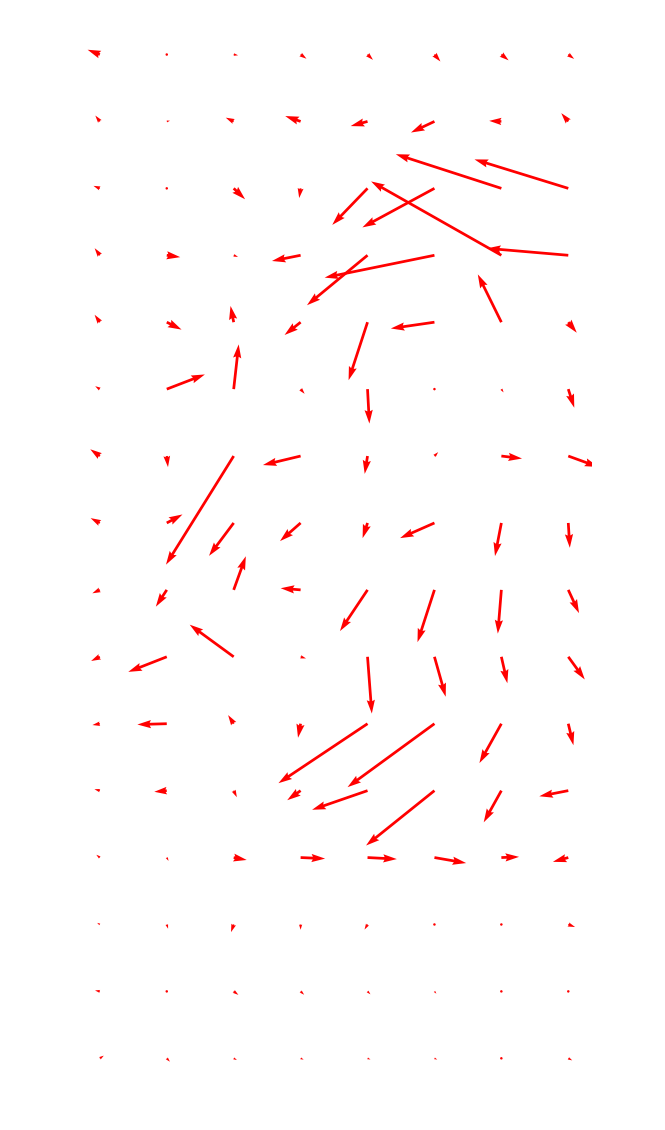}
        \end{subfigure}
        \begin{subfigure}{0.45\textwidth}
            \centering
            \includegraphics[width=\textwidth]{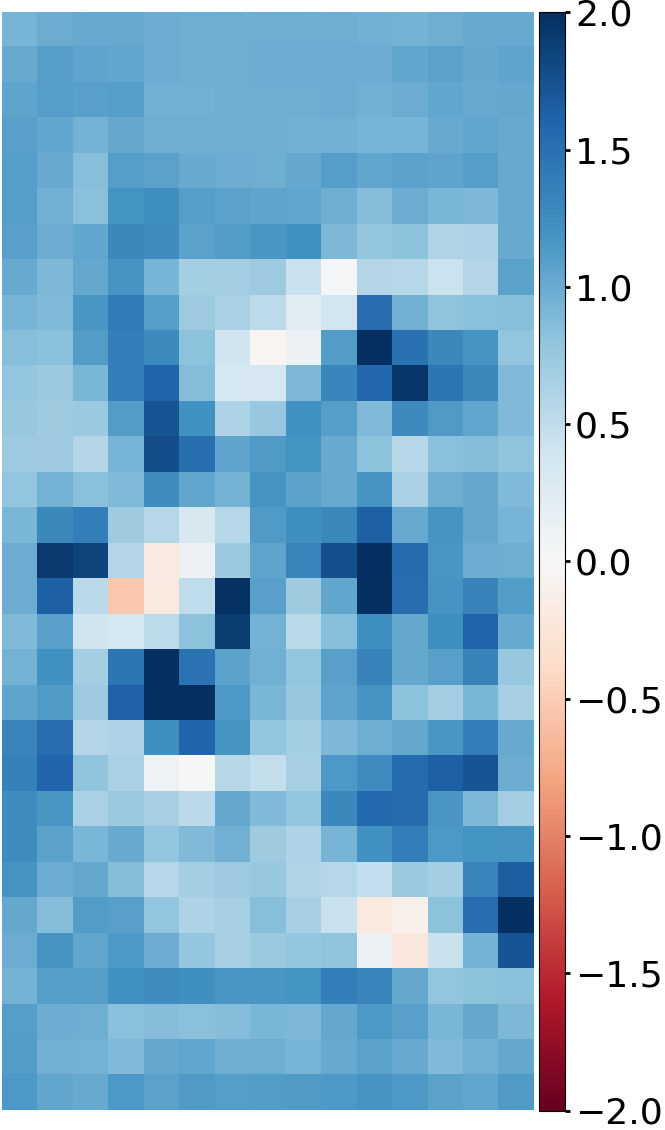}
        \end{subfigure}
        \caption{FeatAlign}
        \label{fig:deformfeatalign}
    \end{subfigure}
        \begin{subfigure}{0.32\textwidth}
        \centering
        \begin{subfigure}{0.45\textwidth}
            \centering
            \includegraphics[width=\textwidth]{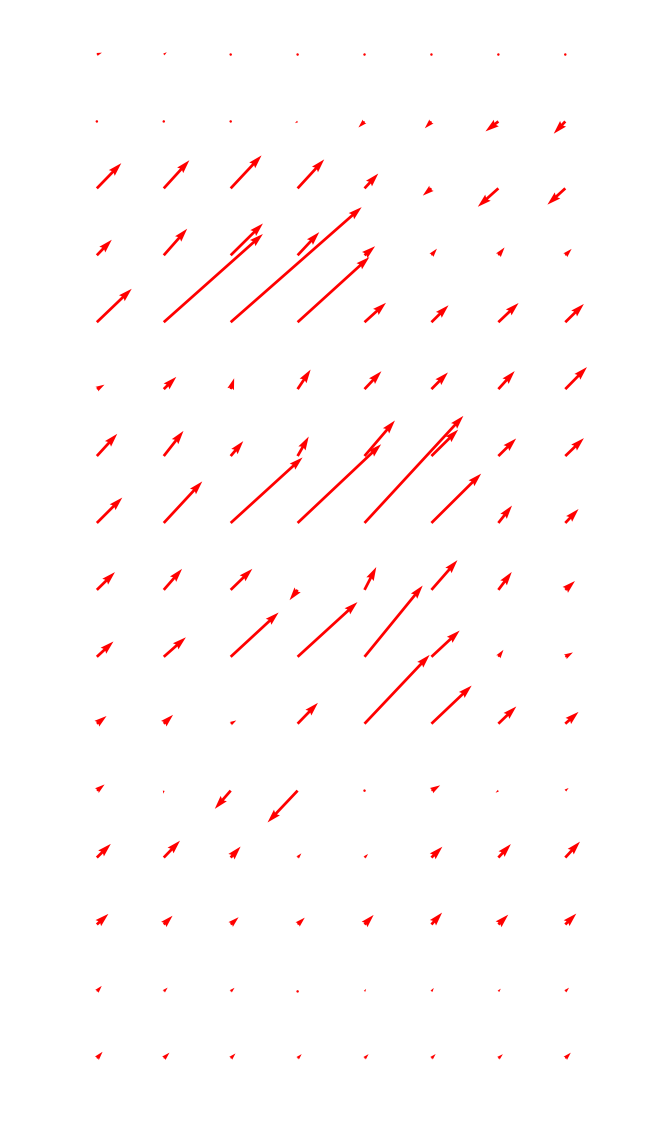}
        \end{subfigure}
        \begin{subfigure}{0.45\textwidth}
            \centering
            \includegraphics[width=\textwidth]{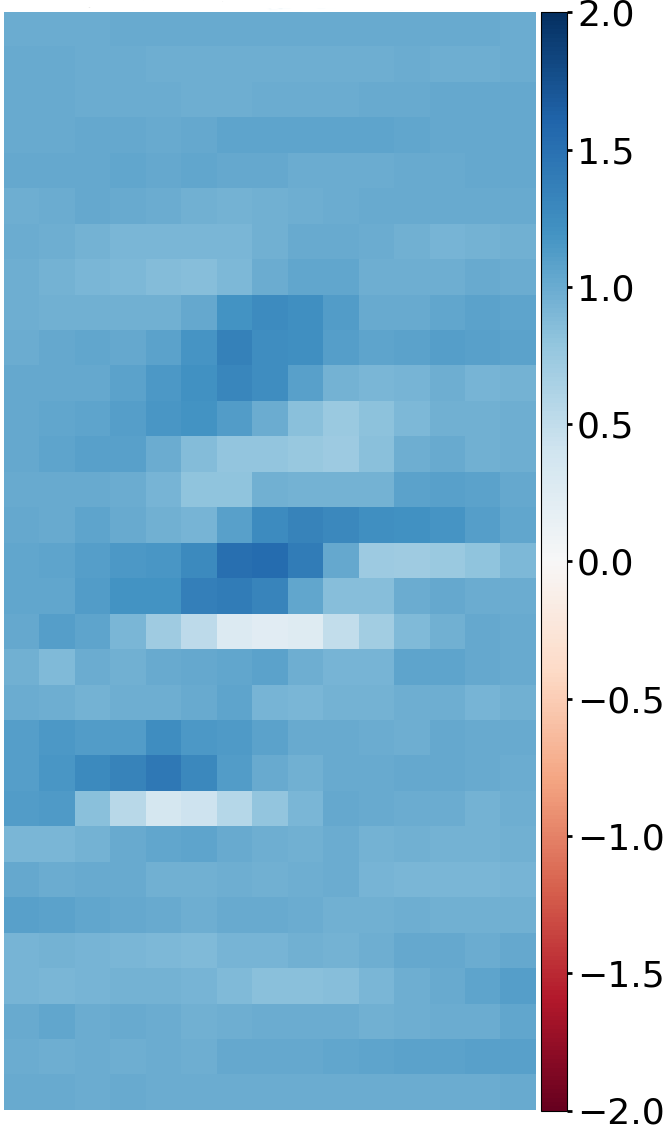}
        \end{subfigure}
        \caption{FeatAlignReg}
        \label{fig:deformfeatalignreg}
    \end{subfigure}
    \begin{subfigure}{0.32\textwidth}
        \centering
        \begin{subfigure}{0.45\textwidth}
            \centering
            \includegraphics[width=\textwidth]{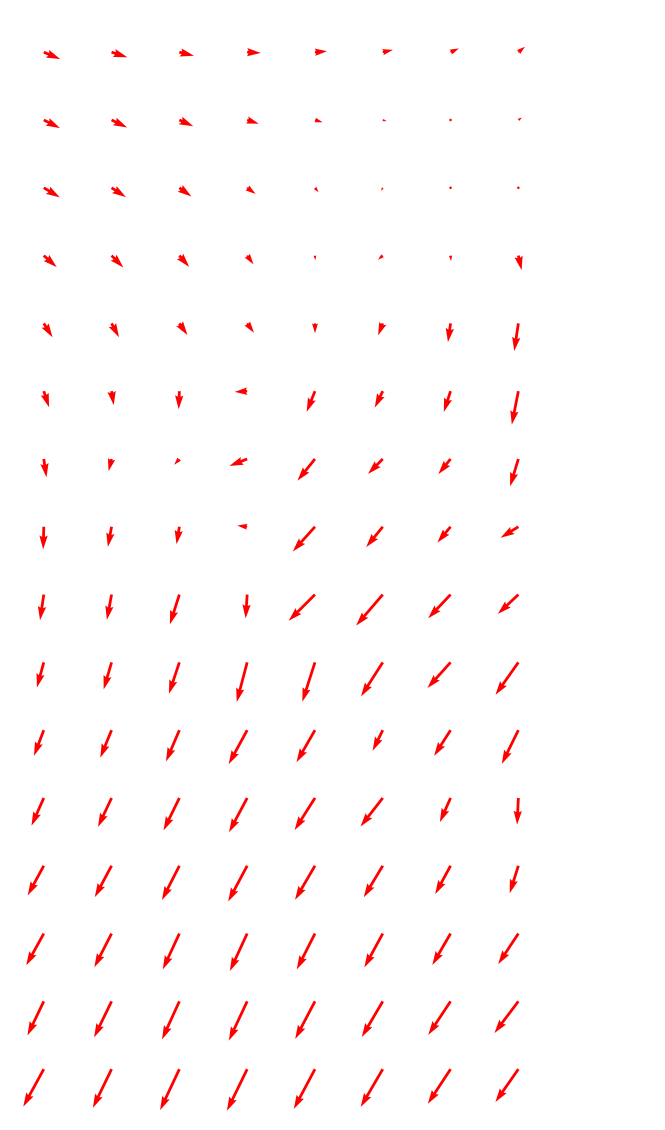}
        \end{subfigure}
        \begin{subfigure}{0.45\textwidth}
            \centering
            \includegraphics[width=\textwidth]{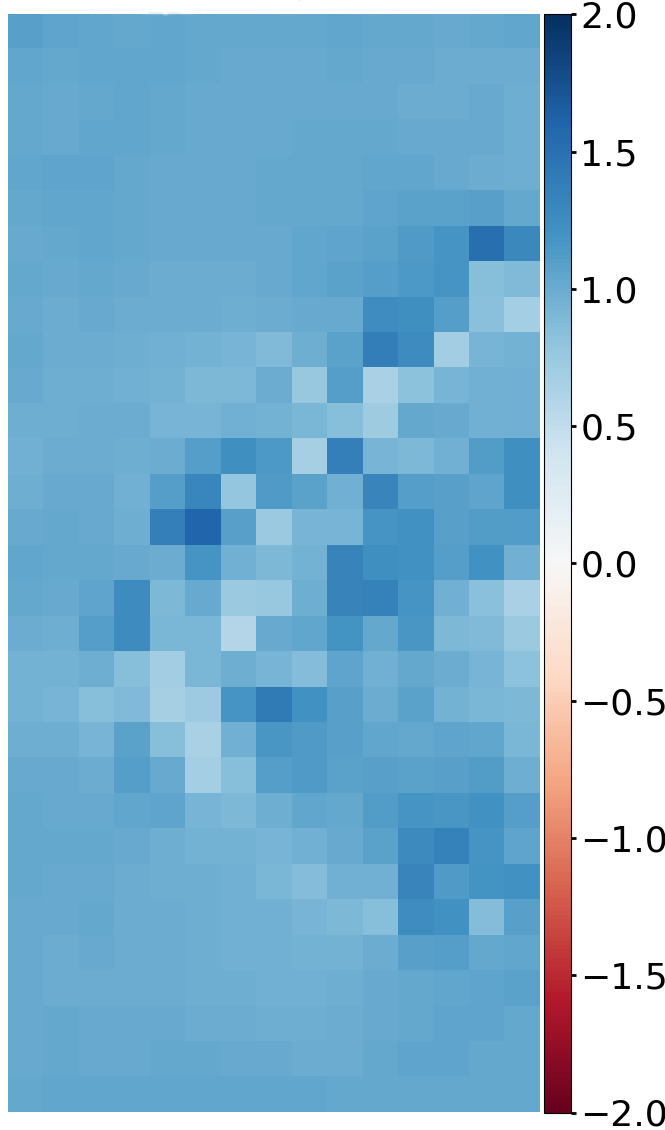}
        \end{subfigure}
        \caption{ImgFeatAlign/ImgAlign}
        \label{fig:deformmammo}
    \end{subfigure}
    \caption{Comparison of deformation field quality. Left: displacement vectors. Right: Jacobian determinants (white/blue: valid, red: invalid).}
    \label{fig:deformationfield}
\end{figure}

While downstream risk prediction performance is one way to assess the quality of the alignment procedures, current approaches within the mammography domain often overlook the quality of the deformation fields, potentially providing an inaccurate picture of how the alignment mechanism benefits the overall performance. We therefore assess the deformation field quality both quantitatively and qualitatively, focusing on displacement vectors and negative Jacobian determinants. FeatAlign methods exhibit significant local deformations  (Fig.~\ref{fig:deformfeatalign}) and the highest percentage of negative Jacobian determinants, reaching $1.6273$. Introducing regularization in FeatAlignReg reduces these local deformations (Fig.~\ref{fig:deformfeatalignreg}) and the percentage of negative Jacobian determinants to $0.0025$, however, this comes at the cost of a slight risk prediction performance drop (see Table~\ref{tab:riskpred}). MammoRegNet, employed in ImgFeatAlign and ImgAlign, further enhances global smoothness by producing a continuous transformation (Fig.~\ref{fig:deformmammo}), resulting in the fewest negative Jacobian determinants at $0.0016$. These findings underscore the benefits of input-level alignment in improving both risk prediction performance and the fidelity of learned deformation fields.

%

\subsubsection{In-Depth Analysis of Weighting and Alignment}
\begin{figure}[tp]
    \centering  
    \begin{subfigure}{0.22\textwidth}
            \centering  
        \includegraphics[height=2.6cm]{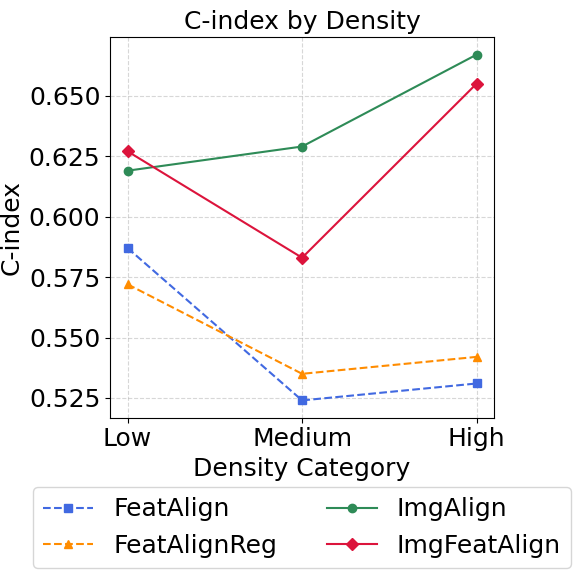}
        \caption{}
        \label{fig:csawdensity}
    \end{subfigure}
    \begin{subfigure}{0.22\textwidth}
        \raggedright 
    \includegraphics[height=2.6cm]{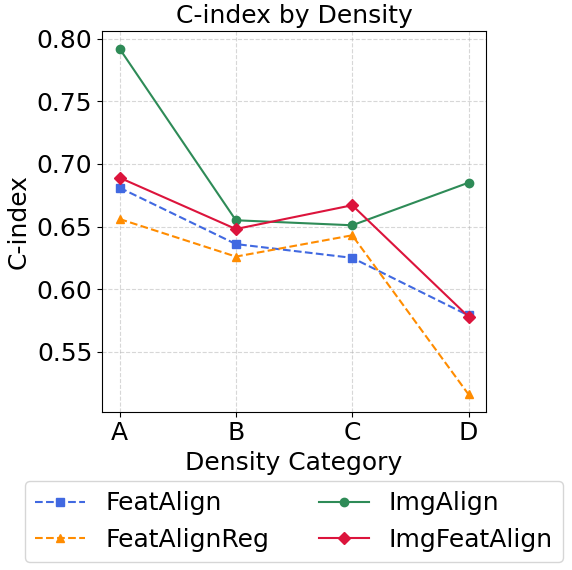}
        \caption{}
      \label{fig:embeddensity}
    \end{subfigure}
    \begin{subfigure}{0.22\textwidth}
            \centering 
    \includegraphics[height=2.6cm]{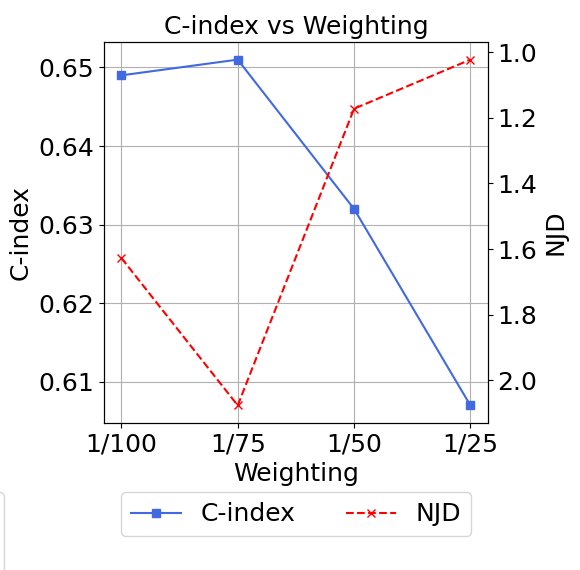}
        \caption{}
        \label{fig:embedcindex}
    \end{subfigure}
     \begin{subfigure}{0.22\textwidth}
         \centering   
    \includegraphics[height=2.6cm]{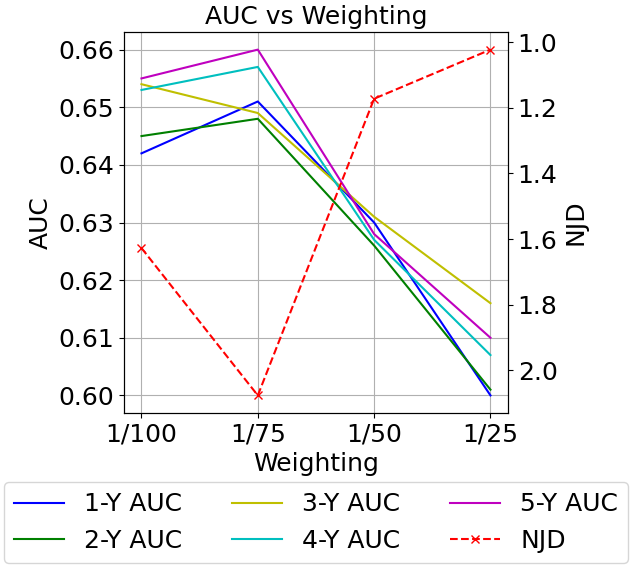}
        \caption{}
                \label{fig:embedauc}
    \end{subfigure}
    \caption{Analysis of Weighting and Alignment: (a) C-index by density category (CSAW-CC), (b) C-index by density category (EMBED), (c) C-index by weighting (EMBED), (d) AUC by weighting (EMBED). }
    \label{fig:wtradeoff}
\end{figure}

 Our findings, along with those in~\cite{Wan_Ordinal_MICCAI2024}, show that feature-level approaches benefit from jointly optimizing risk prediction and alignment. Training them separately reduced risk prediction performance, resulting in a C-index of 62.2, compared to 64.9 (C-index, EMBED dataset). We further explore how different weightings in joint training affect performance and evaluate feature-level and image-level alignment across breast density categories. This distinction is crucial, as higher breast density increases cancer risk and complicates lesion detection. Unlike the EMBED dataset, which includes BI-RADS density categories \cite{birads}, the CSAW-CC dataset provides breast percent density estimates from the LIBRA software \cite{libra}, which we categorized into three equally sized groups, corresponding to low, medium and high density levels. Fig.~\ref{fig:csawdensity} and~\ref{fig:embeddensity}  display the C-index by density category and alignment method for CSAW-CC and EMBED, respectively. ImgAlign performs best, especially in high-density cases, surpassing FeatAlign and FeatAlignReg, highlighting the advantages of image-level alignment for dense breast tissue. 
 Further, to investigate the trade-off between risk prediction performance and deformation field quality, we evaluate the effect of the alignment weight $\alpha$ in Eq.~\ref{lossfeat} on the EMBED dataset in Fig.~\ref{fig:embedcindex}  and~\ref{fig:embedauc}. Increasing the alignment weight improves deformation field quality by reducing negative Jacobian determinants but decreases risk prediction performance emphasizing that these objectives are not necessarily well aligned.

\section{Conclusion and Outlook}
In this study, we provide insights into explicit alignment methods for longitudinal mammograms and their effects on risk prediction performance. Our findings indicate that image-level alignment not only improves the deformation field quality but also enhances risk prediction accuracy and that jointly training feature alignment and risk prediction results in a trade-off between alignment quality and risk prediction performance. Further, to facilitate efficient image-based alignment, we introduce MammoRegNet, the first deep learning-based model for aligning longitudinal mammograms, paving the way towards more accurate risk prediction models. In future work, we aim to further build on this framework that facilitates comparisons between alignment approaches,  extending it to settings where implicit alignment approaches are considered and multiple prior images.

%
%
%
%
\bibliographystyle{splncs04}
\bibliography{main}
\end{document}